\documentstyle[epsf,twoside]{ima}
\contribution{bryan}
\begin{document}

\title{A Hybrid AMR Application for Cosmology and Astrophysics}

\author{GREG L. BRYAN\thanks{
Physics Department, MIT, Cambridge, MA 02139} 
\and MICHAEL L. NORMAN\thanks{
National Center for Supercomputing Applications; and
Astronomy Department, University of Illinois at
Urbana-Champaign, Urbana, IL 61801}
}

\maketitle


\begin{abstract}

We describe an application of Berger and Colella's~\cite{ber89} structured
adaptive mesh refinement algorithm to the field of cosmological
astrophysics.  Simulations in this area must include not just a
gaseous component which follows the hyperbolic equations of
compressible gas dynamics, but also a collisionless component (such as
dark matter or stars) described by the Newtonian dynamical equations.
The two fluids interact via gravity which requires an elliptic solver.
The challenge for AMR is twofold.  First, the collisionless material
is most easily modeled by following trajectories of individual
objects, a method often referred to as an N-body scheme.  The
introduction of particles poses a number of difficulties, both
physical (how do they interact accurately with the fluid in the mesh)
and algorithmic (how to efficiently add a new data structure).
The second challenge is to incorporate the gravitational interaction
between the two components.  We discuss our solutions to both of these
issues and briefly present very encouraging results.

\end{abstract}

\begin{keywords} adaptive mesh refinement, N-body
\end{keywords}


\section{Introduction}

In cosmology and astrophysics, the collapse of objects such as stars
or galaxies from large, diffuse clouds to small, dense cores is both
very important and nearly ubiquitous.  Simulating the gravitational
instability requires resolving a very large dynamic range in three
dimensions, a challenge to traditional static-grid techniques.  This
has lead to the adoption of Lagrangean, particle-based methods such as
Smoothed Particle Hydrodynamics (SPH).  Unfortunately schemes of this
type suffer from a number of drawbacks, such as poor resolution in
shocks and artificially high viscosity.  The structured adaptive mesh
refinement (SAMR) method obviously provides a remedy to this problem
and allows the application of modern, higher-order hydrodynamics
schemes to astrophysics.  However, before such techniques can be
used for many applications in cosmology and astrophysics (and, we
suspect, a number of other disciplines as well), a few problems must
be addressed.

The original SAMR scheme applies mostly to hyperbolic equations
typified by the equations of hydrodynamics.  However, gravity is
described (in the Newtonian limit) by an elliptical equation and,
for many applications, a second, collisionless, fluid is required.
This component may represent stars which, to a high
degree of accuracy, interact only through gravity (in the merger of
two galaxies, only a handful out of $10^{12}$ stars are expected to
physically collide).  Or it may represent the dark matter that
observations of galaxies and clusters of galaxies seem to require.
One way to include this in an SAMR scheme would be to
solve the collisionless Boltzman equation.  This hyperbolic set of
equations follows the evolution of the fluid density in both physical
space and velocity space, together known as phase space.  There are
two obstacles to this approach.  First, for three physical dimensions,
this phase space is six-dimensional, making extreme computational
demands.  Secondly, the equation does not naturally lead to `compact'
solutions, those in which the interesting part of the solution
occupies a small volume in the computational domain.  Therefore, the
SAMR approach is much less appealing for such a fluid.

Fortunately, there is another approach to collisionless systems.  The
N-body method follows trajectories of a representative sample of
individual particles and has proved substantially more efficient than
a direct solution of the Boltzman equation in most situations.  A
substantial amount of effort has been devoted to refining this
approach (see \cite{hoc80} for an excellent summary) and we adopt it
here.  The trajectories are described by the Newtonian dynamical
equations, a set of ordinary differential equations, and this
complicates the SAMR framework somewhat.  The required modifications
are described in the following section and first results are briefly
presented in section~\ref{sec:results}.


\section{Incorporating particles into SAMR}
\label{sec:main}

Here we describe first the additional data structures required by the
introduction of particles and then changes to the AMR method itself.
We assume familiarity with the canonical Berger and Colella SAMR
approach.  A complete description of our implementation will be
made in a future paper.  The hydrodynamics portion of our method uses 
an adaptation of the Piecewise Parabolic Method for cosmological flows
\cite{bn95}.

\subsection{The data structure}

A particle is described by it's position $\vec{x}_p$, velocity
$\vec{v}_p$, and mass $m_p$ (other characteristics are possible but
not necessary).  There is a unique, one-to-one association between a
particle $p$ and a grid $G$ if that particle's position lies within
the grid's boundaries but outside of any finer (child) grid.  We
exploit this association by denoting that grid as the particle's {\it
home} grid and store all such particles in a list along with the rest
of the data connected to that grid. Note that a particle's home grid
may change is it moves, requiring redistribution of particles.  This
disadvantage is offset by a number of 
factors:
\begin{itemize}
 \item decreased search time for particle-grid interactions,
 \item improved data encapsulation, and 
 \item better parallelization characteristics.
\end{itemize}
The third point will be particularly true for a distributed
memory version (which we have not yet implemented).

This association is also very natural from a physical standpoint:
because the particles are indirectly connected to the solution on
their home grid, they tend to share the same time step requirement
(i.e. the maximum time step allowed from accuracy and stability
criteria).

\subsection{The method}

There are two types of method modifications: those related to solving
the new equations and those related to changes in the grid
regeneration.  The first type apply mostly to a single grid and so we
describe them first.  Unless otherwise noted, we discuss the subset of
particles associated with a given (home) grid.

The particle trajectories follow a very simple set of coupled
equations:
\begin{eqnarray}
{d \vec{x}_{p} \over dt} & = &\vec{v}_{p},
          \label{eq:dm_position} \\
{d \vec{v}_{p} \over dt} & = & - \nabla \phi,
          \label{eq:dm_velocity}
\end{eqnarray}
The term on the right-hand side of the second equation is the
gravitational forcing term and it's solution can be found by the
following elliptic equation:
\begin{equation}
\nabla^2 \phi = 4 \pi \rho,
\label{eq:gravity}
\end{equation}
where $\rho$ is the density of both the collisional fluid (grid) and
collisionless fluid (particles).

These equations are finite-differenced and solved with the same
time step as the grid, to reduce bookkeeping.  
Equation~\ref{eq:dm_position} can be 
solved relatively simply as it depends only on quantities local to a
particle, although some care must be taken to correctly time-center
the right-hand side.  The second equation is substantially more
difficult as it involves the solution to an elliptic equation
involving both particle and grid quantities.  The most
straightforward way to solve such an equation is on the mesh, so we
employ the following three-step prescription:
\begin{itemize}
   \item particles $\rightarrow$ grid density field
   \item compute gravitational force on the grid
   \item grid force $\rightarrow$ particles
\end{itemize}
In the first step, a spatially discretized density field is created
out of the particle distribution.  Typically this involves assigning
each particle's mass to the set of 9 or 27 cells nearest to its
location (depending on the desired smoothness of the resulting density
field).  A detailed description of this procedure
can be found in~\cite{hoc80}.

In the second step, eq.~\ref{eq:gravity} is solved with this
density field, which also contains a contribution from the collisional
fluid.  Here we must step back for a moment and recall that this grid,
which we have been discussing in isolation, is actually part of a
hierarchy of patches that, with differing resolutions, covers the
entire computational domain.  A number of techniques have been
suggested to solve such a system (for example, \cite{vil89},
\cite{jes94}, \cite{ann94}), many based on solving the higher levels
first and then interpolating boundary conditions down to the lower
levels.   We have instead adopted a suggestion by Couchman~\cite{cou91}
that utilizes a set of shaped force laws, one for each level of
resolution, that when added together produces the desired forces.  

Once the forces has been computed on the mesh, the particles
associated with that grid must use those forces to update their
velocities via eq.~\ref{eq:dm_velocity}.  This third step is just the
inverse of the first one and the same interpolation technique (using
the nearest 9 or 27 cells) is employed.

The new equations imply a constraint on the grid's time step
since we demand that a particle not move more than some fraction of a
cell width.  This accuracy constraint turns out to be very similar to
the Courant stability criterion.  A slight variation to the original
Berger and Colella control structure is required since the number of
time steps a refined grid takes to catch up to it's parent is no longer
simply the refinement factor.  In fact, strictly speaking, this change
is required for any system of evolution equations where the stability
or accuracy condition depends on the solution, which includes, in
general, fluid dynamics. 

This completes the description of the new equations.  The presence of
particles also affects the grid regeneration process.  Since particles
move, and hence change their home grid, there must be a redistribution
step.  We combine this with the usual SAMR regridding phase and assign
the particles to the newly created grid hierarchy as it is created.
This implies that the regeneration must occur sufficiently often that
particle do not move too far out of the domain of their home grid.


\section{Results}
\label{sec:results}

As one of the final accuracy tests of our combined SAMR-particle
system, we have participated in a joint project to simulate the
formation and evolution of a typical, but challenging, computational
astrophysics target: a cluster of galaxies.  The initial 
conditions, as predicted by a current cosmological theory,
were provided to twelve groups utilizing different numerical
techniques, ranging from SPH to fixed Eulerian grids.  The final
comparison~\cite{sb97} has yet to be completed, however we present
here some results from our SAMR simulation.

\begin{figure}
\epsfxsize=4.5in
\centerline{\epsfbox{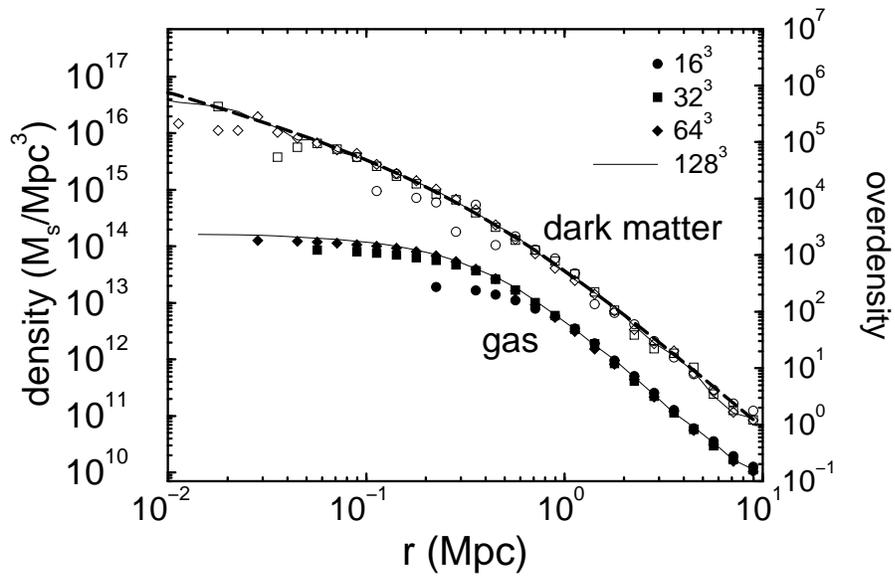}}
\caption{
Dark matter (top curve) and baryonic (bottom curve) radial density
profiles.  Four different runs are shown with varying initial grid
sizes: $16^3$, $32^3$, $64^3$ and the effective $128^3$ run.  The solid
dashed line over the dark matter profile is discussed in the text.
}
\label{fig:amr_profiles}
\end{figure}

The cluster forms from a nearly homogeneous medium and collapses,
under it's own gravity, to a strongly concentrated core.  The
resulting structure is nearly spherical, however, the collapse itself
is strongly asymmetric.  In Figure~\ref{fig:amr_profiles}, we show the
radial density profile of the dark matter (particles) and the gas
(mesh) around the cluster center.   In order to gauge the uncertainty
of this result, we have performed a resolution study, and show the
results with four sets of symbols.  We changed the size of the initial
mesh from 16$^3$ up to 128$^3$ cells.   Although there is some
tendency for the lower resolution versions to produce cores with
somewhat lower central densities, it is clear that the profiles are
nearly converged.  Also plotted, as a dashed line, is the dark matter
profile obtained by other N-body techniques, which agrees well with
our results.  During the evolution, the smallest cell size used by
the SAMR system was 1/8192 of the entire box which provides resolution
comparable to the best SPH codes and far better than a fixed grid
could achieve.


\section{Conclusion}

In this paper we have described one way in which a collisionless fluid
can be modeled, with particles, in an SAMR framework, and have shown
that the resulting system is very successful in modeling at least one
important astrophysical system (reaching an effective resolution
of 8192 in small regions).  We do not
intend to imply that this is the only way -- other variants on these
procedures are possible.  One possibility is that particles could be
split into sub-particles as they enter a refined grid, however, it
is not clear this is a substantial improvement, since the positions and
velocities of the sub-particles are, in general, unknown in advance.  Also,
since the particles don't necessarily exit a refined region at the
same time, recombining them would be problematic.  And, for
gravitational systems at least, the particles are naturally
Lagrangian and hence tend to naturally cluster where the highest
resolution is required, blunting the need to refine them in the
same way as the grid.

Collisionless fluids are a common element in computational
astrophysics, and the N-body technique is the most efficient way to
include them.  However astrophysics is not the only problem domain in
which particles have been used, with examples ranging from plasmas and
vortex simulations to semiconductor devices, and molecular dynamics.
Although structured adaptive mesh refinement techniques may not prove
useful in all these areas, it is likely that the marriage of SAMR and
particles will prove to be more widely applicable, and so we encourage
SAMR software and toolkit developers to build flexible and expandable
systems.

\endcontribution
\end{document}